\documentstyle[12pt]{article}
\begin{document}
\title{Pre-Galaxy Formation:\\
A Non-Linear Analysis of the Evolution of Cosmological Perturbations}
\author{Martin Sch\"on\thanks{e-mail address:
Martin.Schoen@uni-konstanz.de}\\Universit\"at Konstanz\\Fakult\"at
f\"ur Physik\\Postfach 5560 M674, D-78434 Konstanz\\Federal Republic
of Germany}
\date{PASC numbers: 98.80.Dr, 98.60.Ac}
\maketitle
\begin{abstract}
A higher-order analysis of the evolution of cosmological
perturbations in a Friedman universe is given by using the PMF
method. The essence of the PMF approach is to choose a gauge where
all fluctuations of the density, the pressure, and the four-velocity
vanish. In that gauge, even in higher orders, the perturbation field
equations simplify considerably; they can be decoupled and - for
simple equations of state - also be solved analytically. We give the
solution for the dust universe up to third order. Comparison of these
solutions strongly supports the conjecture that in general instable
perturbations grow much faster than they do according to the
first-order analysis. However, perturbations with very large spatial
extension behave differently;  they grow only moderatly. Thus, an
upper boundary of the region of instability seems to exist.
\end{abstract}
\newpage
\section{Introduction}
Due to the non-linearity and complexity of Einstein's field equations
they have been solved analytically so far only for situations which
are characterized by a relatively high symmetry  and homogenity
properties. In the other cases, one has basically two possiblities:
either to solve them numerically, or to perform a perturbation
analysis and to solve the individual orders analytically. Of course,
there are also mixtures of these two possibilities.\\

If we want to explain the origin of galaxies we have to study the
evolution of given small fluctuations in a Friedman universe as
background, and we have to investigate whether these perturbations
are stable or unstable. Moreover, we have to calculate the growth
rates of unstable fluctuations. A complete numerical analysis is not
very useful in this context, because a possible increase does not
guarantee that that perturbation does not decrease somewhen later. On
the other hand, a complete analytical investigation of the evolution
of cosmological perturbations is not possible for reasons of
complexity. Hence, we have to perform some perturbation theory. So
far, that analysis was restricted - also for reasons of complexity -
to the first order; the linearized field equations were studied (see
e.g.\cite{bardeen}, \cite{sakai}, \cite{lifschitz}, \cite{rose},
\cite{paperI}). Only the approach of Ellis et. al. \cite{ellisI},
\cite{ellisII} took into account also the second order. It is
obvious, that such a restriction is too severe. Higher orders become
significant when a perturbation, which is unstable according to the
first-order analysis, is growing. Non-linear effects appearing
thereby could be very important; they could stabilize such a
perturbation, or they could change the growth rates of unstable
perturbations considerably. It is a main aim of this paper to study
what kind of non-linear effects might appear.\\
To this end we have used a method which we have invented in
\cite{paperI}. This so-called PMF method simplified the perturbation
field equations in first order dramatically (see \cite{paperI});
hence, it is reasonable to expect that the same happens for higher
orders, too. The PMF method is based on the gauge freedom. This is to
be understood as follows.\\

Perturbation quantities are constructed by subtracting from the full
quantity at a space-time point $x^{\lambda}$ in the perturbed
universe the background quantity at the corresponding space-time
point $x^{\lambda}$ in the fictitious Friedman universe. The choice
of such a correspondence defines a gauge. We define other gauges by
performing coordinate transformations in the perturbed universe,
keeping the background coordinates fixed. By means of a so-called
gauge condition - a condition concerning some of the perturbation
quantitites - we select a definite set of coordinate systems.
Unfortunately, the perturbation quantities are gauge dependent; thus,
the so-called gauge problem arises: to what extend are the observed
perturbations mere coordinate effects, and to what extend are they
"physical"? Which gauge is suitable for judging stability/instability
of a given fluctuation, how can we find such a gauge which is "as
close to the background as possible"?\\

In the literature one finds some gauges which were proposed in this
context (\cite{sakai}, \cite{lifschitz}, \cite{rose}). The
(first-order) results are systems of coupled differential equations
which can be solved only in special cases. Another treatment in order
to avoid the choice of a gauge was performed by Bardeen
\cite{bardeen}. Out of metric and matter perturbations he constructed
gauge-invariant variables satisfying relatively simple equations
which he could solve explicitly. However, his variables can be used
only partially for a stability analysis: neither does their
limitedness imply the limitedness of the interesting perturbation
quantities like e.g. the density contrast nor does their
unlimitedness cause the latter ones to diverge. Additionally, his
analysis is performed only up to first order.\\
That latter shortcoming is avoided in a recent work of Ellis and
Bruni \cite{ellisI}, \cite{ellisII}. They take into account also the
second order using like Bardeen a gauge-invariant formalism; however,
they do not compare two evolutions (real universe and fictitious
Friedman universe) along the same world line like all approaches up
to now, instead, they rather compare two neighbored world lines
within the same real universe. Their basic quantity is the co-moving
fractional gradient of the energy density orthononal to the fluid
flow. That approach has much in favor but it is quite doubtful
whether it is useful also in higher orders than the second one. The
corresponding equations become quickly very complicated, whereas they
maintain their simple basic structure in the PMF approach, which will
be explained now.\\

The essence of the PMF method is to choose a gauge such that all
fluctuations of matter (i.e., perturbations in density, pressure, and
velocity) vanish and only {\bf p}ure {\bf m}etric {\bf f}luctuations
remain. This can be achieved by choosing suitably the space-like
hypersurfaces ($t$= const.) and by choosing appropriate coordinates
onto them. In first order the perturbation field equations in PMF
gauge simplify considerably, they can be decoupled, and - for simple
equations of state - also be solved \cite{paperI}. Note that the PMF
gauge does not claim to be "as close to the background as possible".
It serves rather for the simplification of the field equations.
Subsequently, one can transform the PMF solution into any desired
gauge suitable for judging stability/instability. In this paper, the
PMF approach shall be employed for a higher-order analysis. It turns
out (see section II) that there is a dramatic simplification also in
higher orders. In the PMF gauge, the perturbation field equations
maintain their simple basic structure which they have in first order.
Therefore, they can be decoupled even in higher orders, and also be
solved analytically if the choosen equation of state is simple
enough.\\

All approaches so far gave no hint that an upper boundary of the
region of instability exists. The first-order result has always been
that fluctuations larger in extension than the Jeans limit
\cite{jeans} are growing eternally (an opposite opinion is supported
in \cite{rose}; but see \cite{paperI}). Thus, one major motivation
for doing higher-orders analysis was the possible perspective that
non-linear effects could imply the existence of such an upper
boundary. Then, fluctuations, whose extension is larger than that
boundary,  would cease to grow or they would grow too slowly with
regard to the generation of large-scale structures of the universe.
Such an existence would be in agreement with astronomical
observations \cite{astronomieI}, \cite{astronomieII},
\cite{astronomieIII}, \cite{gottI}, \cite{gottII}. However, also an
opposite point of view is supported by some astronomers (see e.g.
\cite{deVauc}).\\

The gauge problem itself is of minor interest in this paper. First of
all, we want to analyse higher orders and to study the princple
influence of non-linear effects arising thereby.  Therefore, we will
use a very simple equation of state (dust). In future work the gauge
problem in higher orders as well as more realistic equations of state
shall be investigated.\\
The plan of this paper is as follows. In Sec. II we present the PMF
approach for higher orders and the corresponding perturbation field
equations. These are solved in Sec. III in the case of the dust
universe; a case study, which gives some insight into the
regularities between the solutions of the various orders is
discussed. Sec. IV, which gives the main results of our work, and
which discusses also some other possible fields of application of the
PMF approach, concludes this paper.\\
\\
{\it Notation}: Throughout this paper, we use Weinberg's notation
\cite{weinberg}. Instead of $\partial f/\partial t$ we write
$\dot{f}$; $f_{,1}$ means $\partial f/\partial x$.

\section{Non-linear analysis in the PMF gauge}
For reasons of simplicity, we restrict ourselves to a Friedman
universe with vanishing spatial curvature as background. This is not
too severe a restriction, because in the early universe when galaxies
were formed the condition $\dot{R}^{2}\gg|\kappa|$ ($R$ is the scale
factor of the universe, and $\kappa$ is the curvature parameter,
which can adopt the values $+1$, $0$, or $-1$) was satisfied. We
choose the background coordinates such that the metric components
reads
\begin{equation}
\label{robertsonwalker}
g^{(0)}_{00} = -1, g^{(0)}_{ij} = R(t)^{2}\delta_{ij}
\end{equation}
(all other components vanish). The index 0 refers to the background
($0^{th}$ order).

Let the energy-momentum tensor be of perfect fluid form. We get for
the four-velocity of the background in our Robertson-Walker
coordinates
\begin{equation}
\label{vierer}
U_{(0)}^{\mu} = \delta^{0}_{\mu}.
\end{equation}
Now we consider a perturbation which has two-dimensional symmetry
planes, i.e., in suitable coordinates all perturbation quantities
shall depend only on $x$ ($=x^1$) and $t$ ($=x^0$), but not on $y$
($=x^2$) or $z$ ($=x^3$). Beyond that we demand that $U^{2}_{(n)}$
and $U^{3}_{(n)}$ (n is the order; see (\ref{kansatz})) vanish in
order to exclude rotational perturbations which would disturb the
symmetry.

This symmetry allows the introduction of coordinates such that the
metric adopts the form
\begin{equation}
\label{symmetry}
d\tau^2 = - g_{ab}(x,t)dx^adx^b - f(x,t)(dy^2 + dz^2),
\end{equation}
where $a,b = 0$ oder $1$. Note that $g_{ab}$ and $f$ depend only on
$x$ and $t$.

Next we make the following ansatz
\begin{eqnarray}
\label{kansatz}
\rho & = & \rho_{0} + k \rho_{1} + k^{2} \rho_{2} + k^{3} \rho_{3} +
...,\nonumber\\
p & = & p_{0} + k p_{1} + k^{2} p_{2} + k^{3} p_{3} + ...,\nonumber\\
U^{\mu} & = & U_{(0)}^{\mu} + k U_{(1)}^{\mu} + k^{2} U_{(2)}^{\mu} +
...,\nonumber\\
g_{\mu\nu} & = & g_{\mu\nu}^{(0)} + k g_{\mu\nu}^{(1)} + k^{2}
g_{\mu\nu}^{(2)} + ...,\nonumber\\
\end{eqnarray}
where $k$ is some dimensionless expansion parameter, $\rho$ is the
energy density, $p$ is the pressure, $U^{\mu}$ is the four-velocity
and $g_{\mu\nu}$ is the metric. The index 0 refers to the background,
the index 1 marks the perturbation quantities in first order, the
index 2 those in second order and so on. We have to insert this
ansatz into the field equations and to order them according to powers
of $k$. Subsequently, we have to solve the field equations order by
order. If the sums in (\ref{kansatz}) converge, (\ref{kansatz}) is
the exact solution of the field equations.

The ansatz (\ref{kansatz}) requires some explanations. First of all,
it is not too obvious what is meant by perturbation quantities of
orders higher than the first one. Indeed, one could replace
(\ref{kansatz}) by an alternative ansatz which just contains
perturbation quantitities up to first order. $k\rho_{1}$ then simply
means the difference between the total density in the real universe
and the density $\rho_{0}$ in the fictitious background universe. But
contrary to the linear analysis performed in \cite{paperI} we had, in
this case, also to take into account terms proportional to $k^n$
($n\geq2$) as e.g. the term $g_{\mu\nu}^{(1)}g_{\tau\lambda}^{(1)}$
in the field equations. Then, we could not expect that the field
equations are satisfied order by order; merely the sum of all orders
from the first one to the last considered one would be satisfied. We
do not want to choose this approach here for in that case we had to
solve practically the full field equations; just the zeroth order
would be separeted. This would not be really a perturbation analysis.

Instead of this we proceed as follows. In order to satisfy the field
equations order by order we add in our ansatz a correction quantity
order by order, which is chosen such that these equations hold. Doing
so we arrive at the ansatz (\ref{kansatz}) and it is guaranteed that
in each order $n$ the field equations are differential equations
which are linear in the unknown correction quantitites $\rho_{n}$,
$p_{n}$, etc., (but not in perturbation quantities of lower orders
which are already known by solving the field equations in these lower
orders; hence, they are no longer unknown with regard to the $n^{th}$
order). Note that either $k$ can be considered as small compared with
1, or we can set $k=1$; then, the perturbation quantities of
$(n+1)^{th}$ order like $\rho_{(n+1)}$, $U^{\mu}_{(n+1)}$, etc should
be small compared with the corresponding ones of the $n^{th}$ order.

Now we study the influence of infinitesimal coordinate
transformations
\begin{equation}
\label{trafo}
x'^{\mu} = x^{\mu} - k\epsilon_{(1)}^{\mu}(x^{\lambda}) -
k^{2}\epsilon_{(2)}^{\mu}(x^{\lambda}) -
k^{3}\epsilon_{(3)}^{\mu}(x^{\lambda}) - ...
\end{equation}
in the real perturbed universe. These coordinate transformations
change the correspondence between points in the background universe
and points in the physical space-time and are, hence, gauge
transformations. Note that the "philosophy" of the ansatz
(\ref{trafo}) is the same as that of (\ref{kansatz}): if we would
stop the sum in (\ref{trafo}) at the term
$\epsilon_{(1)}^{\mu}(x^{\lambda})$ we would get very complicated
non-linear gauge transformation laws for the quantities apearing in
(\ref{kansatz}) when we would go beyond the first order. But with
(\ref{trafo}) we get in each order $n$ transformation laws which are
linear in the unknown function $\epsilon_{(n)}^{\mu}$. We give them
here explicitly up to the second order:\\
\\
First order:\\
\begin{eqnarray}
\label{rhotrafo}
\rho_{1}' &=& \rho_{1} + \dot{\rho}_{0}\epsilon_{(1)}^{t},\\
\label{ptrafo}
p_{1}' &=& p_{1} + \dot{p}_{0}\epsilon_{(1)}^{t},\\
\label{Utrafo}
U_{(1)}'^{\mu} &=& U_{(1)}^{\mu} - \dot{\epsilon}_{(1)}^{\mu},\\
\label{gtrafo}
g'_{\mu \nu}^{(1)} &=& g_{\mu \nu}^{(1)} + \epsilon_{(1)\mu};\nu +
\epsilon_{(1)\nu};\mu.
\end{eqnarray}
(cf. Eqs. (2.4), (2.5), (2.7), (2.9) in \cite{paperI})
\\
Second order:\\
\begin{eqnarray}
\label{rhotrafo2}
\rho'_{2} &=& \rho_{2} + \dot{\rho}_{0}\epsilon_{(2)}^{t} +
\frac{\partial\rho_{1}}{\partial x^{\mu}}\epsilon_{(1)}^{\mu} +
\frac{\ddot{\rho_{0}}}{2}(\epsilon_{(1)}^{t})^{2},\\
\label{ptrafo2}
p_{2}' &=& p_{2} + \dot{p}_{0}\epsilon_{(2)}^{t} + \frac{\partial
p_{1}}{\partial x^{\mu}}\epsilon_{(1)}^{\mu} +
\frac{\ddot{p_{0}}}{2}(\epsilon_{(1)}^{t})^{2},\\
\label{Utrafo2}
U_{(2)}'^{\mu} &=& U_{(2)}^{\mu} - \dot{\epsilon}_{(2)}^{\mu} +
\frac{\partial U_{(1)}^{\mu}}{\partial x^{\mu}}\epsilon_{(1)}^{\mu} -
\frac{\partial \epsilon_{(1)}^{\mu}}{\partial
x^{\nu}}U_{(1)}^{\nu},\\
\label{gtrafo2}
g'_{\mu\nu}^{(2)} &=& g_{\mu\nu}^{(2)} + \epsilon_{(2)\mu};\nu +
\epsilon_{(2)\nu};\mu - g^{(0)}_{\lambda\kappa}\frac{\partial
\epsilon_{(1)}^{\lambda}}{\partial x^{\mu}}\frac{\partial
\epsilon_{(1)}^{\kappa}}{\partial x^{\nu}}\nonumber\\
 &+& g'_{\mu\lambda}^{(1)}\frac{\partial
\epsilon_{(1)}^{\lambda}}{\partial x^{\nu}}
+g'_{\nu\lambda}^{(1)}\frac{\partial
\epsilon_{(1)}^{\lambda}}{\partial x^{\mu}} + \frac{\partial
g^{(0)}_{\mu\lambda}}{\partial x^{\kappa}}\frac{\partial
\epsilon_{(1)}^{\lambda}}{\partial
x^{\nu}}\epsilon_{(1)}^{\kappa}\nonumber\\
 &+& \frac{\partial g^{(0)}_{\nu\lambda}}{\partial
x^{\kappa}}\frac{\partial \epsilon_{(1)}^{\lambda}}{\partial
x^{\mu}}\epsilon_{(1)}^{\kappa} + \frac{g'_{\mu\nu}^{(1)}}{\partial
x^{\lambda}}\epsilon_{(1)}^{\lambda} -
\frac{1}{2}\frac{\partial^{2}g^{(0)}_{\mu\nu}}{\partial
x^{\lambda}x^{\kappa}}\epsilon_{(1)}^{\lambda}\epsilon_{(1)}^{\kappa}.

\end{eqnarray}
According to (\ref{rhotrafo}) and (\ref{ptrafo}) as well as
(\ref{rhotrafo2}) and (\ref{ptrafo2}) it is possible to transform
$\rho'_{1}$, $p'_{1}$, $\rho'_{2}$ and $p'_{2}$ simultaneously to
zero by a suitable choice of $\epsilon_{(1)}^{t}$ and
$\epsilon_{(2)}^{t}$ provided that $\dot{\rho_{0}}$ is different from
zero (which is satisfied in an expanding universe) and provided that
the following equation holds for $i = 0,1$, and $2$
\begin{equation}
\label{PMFbedingung}
p_{i} = \omega\rho_{i},
\end{equation}
where $\omega$ is spatially and temporally constant.
(\ref{PMFbedingung}) is certainly satisfied, even for all natural
numbers $i$, if the equation of state $p = \omega\rho$ holds (insert
the ansatz (\ref{kansatz}) and decompose $p$ according to powers of
$k$ such that this equation is satisfied order by order).
Analogeously, $U'_{(1)}^{1}$ and $U'_{(2)}^{1}$ can be transformed
away by a suitable choice of $\epsilon_{(1)}^{1}$ and
$\epsilon_{(2)}^{1}$. All other spatial components of $U^{i}$ also
vanish due to our symmetry assumption. (Note that $U'_{(i)}^{0}$ is
not an independent variable because it is related to $g^{(i)}_{00}$
by the norm conservation). We have now all features of the PMF-gauge
present: fluctuations of the density, of the pressure, and of the
spatial components of the four-velocity vanish simultaneously (in
first and in second order). One can show easily that this can be
achieved also in higher orders. Namely, the gauge transformation laws
for the perturbation quantities have in each order $n$ the following
general structure:\\
\begin{eqnarray}
\label{rhotrafoN}
\rho'_{n} &=& \rho_{n} + \dot{\rho}_{0}\epsilon_{(n)}^{t} +
f_{\rho}(\rho_{i},\epsilon_{(j)}),\\
\label{ptrafoN}
p'_{n} &=& p_{n} + \dot{p}_{0}\epsilon_{(n)}^{t} +
f_{p}(p_{i},\epsilon_{(j)}),\\
\label{UtrafoN}
U'_{(n)}^{\mu} &=& U_{(n)}^{\mu} - \dot{\epsilon}_{(n)}^{\mu} +
f_{U}(U_{(i)}^{\mu},\epsilon_{(j)}),
\end{eqnarray}
where $f_{\rho}(\rho_{i},\epsilon_{(j)})$ is some function of
$\rho_{i}$ and $\epsilon_{(j)}$ and of their derivatives. The
subscripts $i$ and $j$ are smaller than the considered order $n$. The
other functions appearing in these transformation laws must be
interpreted in the same way. As can be read off from
(\ref{rhotrafoN}), (\ref{ptrafoN}), and (\ref{UtrafoN}) it is
possible to transform $\rho_{(n)}'$, $p_{(n)}'$, and $U'_{(n)}^{1}$
simultaneously to zero by a suitable choice of $\epsilon_{(n)}^{t}$
and $\epsilon_{(n)}^{1}$ provided that (\ref{PMFbedingung}) holds for
all natural numbers $i$.\\

Hence, if the condition (\ref{PMFbedingung}) is satisfied we can
achieve:
\begin{eqnarray}
\label{PMF}
\rho &=& \rho_{0},\nonumber\\
p &=& p_{0},\nonumber\\
U^{i} &=& U^{i}_{(0)}, i = 1,2,3 ,
\end{eqnarray}
But (\ref{PMF}) characterizes exactly the PMF gauge (see
\cite{paperI}). From now on we assume that (\ref{PMFbedingung}) is
satisfied and we are going to solve the field equations in PMF gauge
characterised by (\ref{PMF}). Note that because of (\ref{symmetry})
$g_{22}^{(n)} = g_{33}^{(n)}$ holds for all orders and that besides
these two metric components only $g_{00}^{(n)}$, $g_{10}^{(n)}$ and
$g_{11}^{(n)}$ can be non-vanishing. Hence, our the field equations
contain just 4 independent components. Like in \cite{paperI} we are
using instead of the components "11" and "22" the energy-momentum
conservation. Clearly, for reasons of checking we have inserted the
solution obtained thereby into all components of the field
equations.\\
After a very lengthy but straightforward calculation we get the field
equations up to third order in the following form where the different
orders are already entangled from each other:\\
\begin{eqnarray}
\label{eqs,n}
\mbox{00-component:}&&\nonumber\\
{}\nonumber\\
\left[8\pi G\rho_{0} - 6(\frac{\dot{R}}{R})^{2}\right]g_{00}^{(n)} +
2\frac{\dot{R}}{R^{3}}g_{10,1}^{(n)} + \frac{g_{22,11}^{(n)}}{R^{4}}
&=& S_{00}^{(n-1)}\nonumber\\
{}\nonumber\\
\mbox{10-component:}&&\nonumber\\
{}\nonumber\\
\left[3(\frac{\dot{R}}{R})^{2} + 8\pi
Gp_{0}\right]\frac{g_{10}^{(n)}}{R^{2}} -
\frac{\dot{R}}{R^{3}}g_{00,1}^{(n)}
-\frac{\dot{g}_{22,1}^{(n)}}{R^{4}}
+\frac{2\dot{R}}{R^{5}}g_{22,1}^{(n)} &=& S_{10}^{(n-1)}\nonumber\\
{}\nonumber\\
\mbox{energy conservation:}&&\nonumber\\
{}\nonumber\\
\frac{\dot{g}_{11}^{(n)}}{R^{2}} + 2\frac{\dot{g}_{22}^{(n)}}{R^{2}}
- \frac{2\dot{R}}{R^{3}}(g_{11}^{(n)} + 2g_{22}^{(n)}) &=&
S_{ec}^{(n-1)}\nonumber\\
{}\nonumber\\
\mbox{momentum conservation:}&&\nonumber\\
{}\nonumber\\
\frac{\dot{g}_{10}^{(n)}}{R^{2}} - \frac{g_{00,1}}{2R^{2}} + \left[
\frac{\dot{p_{0}} }{p_{0} +
\rho_{0}}\right]\frac{g_{10}^{(n)}}{R^{2}} &=& S_{mc}^{(n-1)}
\end{eqnarray}
The source terms $S_{\mu\nu}^{(n-1)}$ are sums of products whose
factors are solutions (or their derivations) of the field equations
of lower orders than $n$ (i.e., maximally of $(n-1)^{th}$ order). We
have calculated them for $n = 1,2,3$. We give the full form of the
equations (\ref{eqs,n}) in the appendix, but only up to second order
since for $n=3$ they are horribly long. However, after inserting the
solutions for lower orders the source terms simplify quite a lot and
can be handled e.g. in the case of the dust universe (see next
section) quite easily. For $n=1$ all source terms vanish; the system
(\ref{eqs,n}) is then already familiar to us from the first order
analysis (compare \cite{paperI}, (3.17) - (3.20), and - considering
the momentum conservation equation - note that the relation
$\dot{\rho}_{0} + 3(p_{0} + \rho_{0})\dot{R}/R = 0$ (that is the
energy conservation in $0^{th}$ order) holds). Because of that
striking regularity it can be suspected that (\ref{eqs,n}) is valid
for all orders $n$; but so far this conjecture has been proven by us
(by directly working out the full form of (\ref{eqs,n})) only for
$n=1,2$, and $3$.\\

Eq.(\ref{eqs,n}) can be decoupled like in the first order (see
\cite{paperI}) also in higher orders - because of its relativly
simple structure. Moreover, for simple equations of state, it can be
solved completely in a purely analytical manner. We did this for dust
($p=0$) and for radiation ($p=\rho/3)$ but in this paper we merely
discuss the dust solution. This will be done in the next section.
\section{The dust universe}
The dust universe is characterized by a vanishing pressure. Although
such an equation of state is not very realistic concerning the
evolution of galaxies, the following discussion shows important
features of the application of the PMF method to higher orders. In
particular, we will find new non-linear effects. The following study
of the dust universe can be considered also as some kind of training
for handling the full (i.e., not restricted to the first order) PMF
method.
The Friedman equations imply
\begin{equation}
\label{staubkosmos}
R(t) = Kt^{2/3}, K = const.,\rho_{0}=(6\pi G t^{2})^{-1}, p_{0} = 0.
\end{equation}
If we insert this into our system of eqs.(\ref{eqs,n}) and decouple
it in the same manner as we did it in first order in \cite{paperI},
we obtain the following results:\\
\\
{\bf First Order:}\\
\begin{eqnarray}
\label{g1,10}
g^{(1)}_{10}&=&\left( { A_{2}} + { A_{1}}\,{t^{{5\over 3}}} \right)
\,
  \cos (q\,x)\\
\label{g1,00}
g^{(1)}_{00}&=&{{10\,{ A_{1}}\,{t^{{2\over 3}}}\,\sin (q\,x)}\over
   {3\,q}}\\
\label{g1,11}
g^{(1)}_{11}&=&\left( {{8\,{ A_{2}}\,{K^2}\,{t^{{1\over 3}}}}\over
      {3\,q}} + {{80\,{ A_{1}}\,{K^4}\,
        {t^{{4\over 3}}}}\over {9\,{q^3}}} +
    {{8\,{ A_{1}}\,{K^2}\,{t^2}}\over {3\,q}} +
    {K^2}\,{t^{{4\over 3}}}\,\tau  \right) \,\sin (q\,x)\\
\label{g1,22}
g^{(1)}_{22}&=&\left( {{-4\,{ A_{2}}\,{K^2}\,{t^{{1\over 3}}}}\over
      {3\,q}} - {{40\,{ A_{1}}\,{K^4}\,
        {t^{{4\over 3}}}}\over {9\,{q^3}}} -
    {{4\,{ A_{1}}\,{K^2}\,{t^2}}\over {3\,q}} \right) \,
  \sin (q\,x)
\end{eqnarray}
where q is the wave number (for reasons of simplicity we choose
solutions proportional to $exp(iqx)$); $A_{1}$ and $A_{2}$ are
constants which are related in an unique way to the "history" of the
universe in question (i.e. the state of the perturbed universe, see
\cite{paperI}); $\tau$, however, merely reflects the remaining
freedom of performing  transformations within the PMF gauge, and has,
hence, no deeper meaning (see again \cite{paperI}). Note that the
constants $A_{1}$ and $A_{2}$ are defined slightly different as those
in \cite{paperI}.\\
\\
{\bf Second Order:}\\

The decoupling of (\ref{eqs,n}) in second order is performed in the
same way as in first order. The second order equations are also
analytically relatively easy to solve, because the source terms
$S_{\mu\nu}^{(1)}$ are - after inserting the first order dust
solutions (\ref{g1,10}), (\ref{g1,00}), (\ref{g1,11}) and
(\ref{g1,22}) - just finite sums of powers of $t$. We obtain:\\
\begin{eqnarray}
\label{g2,10}
g^{(2)}_{10}&=&\left[ - {{{{{ A_{2}}}^2}}\over {2\,q\,t}} -
  {{7\,{ A_{1}}\,{ A_{2}}\,{t^{{2\over 3}}}}\over
    {2\,q}}-
  {{3\,{{{ A_{1}}}^2}\,{t^{{7\over 3}}}}\over q}+{ B_{2}} + {{3\,{
B_{1}}\,{t^{{5\over 3}}}}\over 5}\right]\,\sin(2q\,x)\\
\label{g2,00}
g^{(2)}_{00}&=&\left[-{{{{{ A_{2}}}^2}}\over {2\,{q^2}\,{t^2}}} +
  {{16\,{ A_{1}}\,{ A_{2}}}\over
    {9\,{q^2}\,{t^{{1\over 3}}}}} +
  {{58\,{{{ A_{1}}}^2}\,{t^{{4\over 3}}}}\over {9\,{q^2}}}-
  {{{ B_{1}}\,{t^{{2\over 3}}}}\over
q}\right]\,\cos(2q\,x)\nonumber\\
&+&\,\left[{{{{{ A_{2}}}^2}}\over {2\,{q^2}\,{t^2}}} -
  {{16\,{ A_{1}}\,{ A_{2}}}\over
    {9\,{q^2}\,{t^{{1\over 3}}}}} -
  {{50\,{{{ A_{1}}}^2}\,{K^2}\,{t^{{2\over 3}}}}\over
    {27\,{q^4}}} - {{58\,{{{ A_{1}}}^2}\,
      {t^{{4\over 3}}}}\over {9\,{q^2}}}\right]\\
\label{g2,11}
g^{(2)}_{11}&=&\left[{{-{{{ A_{2}}}^2}}\over 2} -
  {{16\,{{{ A_{2}}}^2}\,{K^2}}\over
    {9\,{q^2}\,{t^{{2\over 3}}}}} -
  {{440\,{ A_{1}}\,{ A_{2}}\,{K^4}\,{t^{{1\over 3}}}}\over
      {27\,{q^4}}} - {{4\,{ B_{2}}\,{K^2}\,
      {t^{{1\over 3}}}}\over {3\,q}} -
  {{4\,{ A_{1}}\,{ A_{2}}\,{K^2}\,t}\over {3\,{q^2}}} +
  { B_{3}}\,{t^{{4\over 3}}}\right.\nonumber\\
&-&\,\left.{ A_{1}}\,{ A_{2}}\,{t^{{5\over 3}}} -
  {{440\,{{{ A_{1}}}^2}\,{K^4}\,{t^2}}\over
    {27\,{q^4}}} - {{4\,{ B_{1}}\,{K^2}\,{t^2}}\over
    {5\,q}} + {{4\,{{{ A_{1}}}^2}\,{K^2}\,
      {t^{{8\over 3}}}}\over {9\,{q^2}}} -
  {{{{{ A_{1}}}^2}\,{t^{{{10}\over 3}}}}\over
2}\right]\,\cos(2q\,x)\nonumber\\
&-&\,{{{{{ A_{2}}}^2}}\over 2} +
  {{16\,{{{ A_{2}}}^2}\,{K^2}}\over
    {9\,{q^2}\,{t^{{2\over 3}}}}} -
  { F_{3}}\,{K^2}\,{t^{{1\over 3}}} +
  {{4\,{ A_{1}}\,{ A_{2}}\,{K^2}\,t}\over {3\,{q^2}}} +
  { F_{4}}\,{K^2}\,{t^{{4\over 3}}}\nonumber\\
&-&\,
  { A_{1}}\,{ A_{2}}\,{t^{{5\over 3}}} +
  {{160\,{{{ A_{1}}}^2}\,{K^4}\,{t^2}}\over
    {27\,{q^4}}} - {{4\,{{{ A_{1}}}^2}\,{K^2}\,
      {t^{{8\over 3}}}}\over {9\,{q^2}}} -
  {{{{{ A_{1}}}^2}\,{t^{{{10}\over 3}}}}\over 2}\\
\label{g2,22}
g^{(2)}_{22}&=&\left[{{-4\,{{{ A_{2}}}^2}\,{K^2}}\over
    {9\,{q^2}\,{t^{{2\over 3}}}}} -
  {{20\,{ A_{1}}\,{ A_{2}}\,{K^4}\,{t^{{1\over 3}}}}\over
    {27\,{q^4}}} + {{2\,{ B_{2}}\,{K^2}\,
      {t^{{1\over 3}}}}\over {3\,q}} -
  {{2\,{ A_{1}}\,{ A_{2}}\,{K^2}\,t}\over {{q^2}}}\right.\nonumber\\
&+&\,
  {{350\,{{{ A_{1}}}^2}\,{K^6}\,{t^{{4\over 3}}}}\over
    {81\,{q^6}}} + {{{ B_{1}}\,{K^4}\,
      {t^{{4\over 3}}}}\over {3\,{q^3}}} -
  {{20\,{{{ A_{1}}}^2}\,{K^4}\,{t^2}}\over {27\,{q^4}}} +
  {{2\,{ B_{1}}\,{K^2}\,{t^2}}\over {5\,q}}\nonumber\\
&-&\,\left.
  {{14\,{{{ A_{1}}}^2}\,{K^2}\,{t^{{8\over 3}}}}\over
    {9\,{q^2}}}\right]\,\cos(2q\,x)\nonumber\\
&+&\,{{4\,{{{ A_{2}}}^2}\,{K^2}}\over
    {9\,{q^2}\,{t^{{2\over 3}}}}} -
  { F_{1}}\,{K^2}\,{t^{{1\over 3}}} +
  {{2\,{ A_{1}}\,{ A_{2}}\,{K^2}\,t}\over {{q^2}}} +
  { F_{2}}\,{K^2}\,{t^{{4\over 3}}}\nonumber\\
&+&\,
  {{160\,{{{ A_{1}}}^2}\,{K^4}\,{t^2}}\over
    {27\,{q^4}}} + {{14\,{{{ A_{1}}}^2}\,{K^2}\,
      {t^{{8\over 3}}}}\over {9\,{q^2}}}
\end{eqnarray}
The new parameters $B_{i}$ and $F_{i}$ arise by solving the
homogeneous part of the system of differential equations
(\ref{eqs,n}); the terms which contain the parameters $A_{i}$, which
are already known from our first order analysis, are generated as
special inhomogeneous solutions by the source terms
$S_{\mu\nu}^{(1)}$. The integration constant $\tau$, which can be
transformed to zero within the PMF gauge, has been omitted. While the
$B_{i}$ can be chosen freely, the parameters $F_{i}$ have to satisfy
the following equation:\\
\begin{equation}
\label{bedingung}
160\,{ A_{1}}\,{ A_{2}}\,{K^2} + 18\,{ F_{1}}\,{q^4} +
  9\,{ F_{3}}\,{q^4} = 0.
\end{equation}
Note that the transition from the first order to the second one
causes a doubling of the wave number $q$! Additionally, terms which
are spatially constant arise.\\
\\
{\bf Third Order:}\\

Again, the source terms turn out to be - after inserting the just
obtained dust solutions in first and second order - finite sums of
powers of $t$. Hence, (\ref{eqs,n}) is also in third order easy to
solve. We get:\\
\begin{eqnarray}
\label{g3,10}
g^{(3)}_{10}&=&\left[{ C_{2}} + {{{{{ A_{2}}}^3}}\over
{8\,{q^2}\,{t^2}}} +
  {{5\,{ A_{1}}\,{{{ A_{2}}}^2}\,{K^2}}\over
    {2\,{q^4}\,t}} + {{3\,{ A_{2}}\,{ B_{2}}}\over
    {4\,q\,t}} - {{109\,{ A_{1}}\,{{{ A_{2}}}^2}}\over
    {36\,{q^2}\,{t^{{1\over 3}}}}} +
  {{40\,{{{ A_{1}}}^2}\,{ A_{2}}\,{K^2}\,
      {t^{{2\over 3}}}}\over {3\,{q^4}}}\right.\nonumber\\
 &+&\,
  {{29\,{ A_{2}}\,{ B_{1}}\,{t^{{2\over 3}}}}\over
    {20\,q}} + {{17\,{ A_{1}}\,{ B_{2}}\,
      {t^{{2\over 3}}}}\over {6\,q}} -
  {{763\,{{{ A_{1}}}^2}\,{ A_{2}}\,{t^{{4\over 3}}}}\over
    {72\,{q^2}}} + {{3\,{ C_{1}}\,{t^{{5\over 3}}}}\over
    5} + {{40\,{{{ A_{1}}}^3}\,{K^2}\,
      {t^{{7\over 3}}}}\over {21\,{q^4}}}\nonumber\\
&+&\,
  \left.{{27\,{ A_{1}}\,{ B_{1}}\,{t^{{7\over 3}}}}\over
    {10\,q}} - {{67\,{{{ A_{1}}}^3}\,{t^3}}\over
    {9\,{q^2}}}\right]\,\cos(3q\,x)\nonumber\\
&+&\,\left[{ C_{4}} - {{{{{ A_{2}}}^3}}\over {8\,{q^2}\,{t^2}}} -
  {{{ A_{2}}\,{ F_{1}}}\over {4\,t}} +
  {{{ A_{2}}\,{ F_{3}}}\over {4\,t}} +
  {{25\,{ A_{1}}\,{{{ A_{2}}}^2}\,{K^2}}\over
    {18\,{q^4}\,t}} - {{{ A_{2}}\,{ B_{2}}}\over
    {4\,q\,t}} + {{109\,{ A_{1}}\,{{{ A_{2}}}^2}}\over
    {36\,{q^2}\,{t^{{1\over 3}}}}} -
  {{3\,{ A_{1}}\,{ F_{1}}\,{t^{{2\over 3}}}}\over
2}\right.\nonumber\\
 &+&\,
  {{{ A_{1}}\,{ F_{3}}\,{t^{{2\over 3}}}}\over 4} +
  {{25\,{{{ A_{1}}}^2}\,{ A_{2}}\,{K^2}\,
      {t^{{2\over 3}}}}\over {27\,{q^4}}} -
  {{3\,{ A_{2}}\,{ B_{1}}\,{t^{{2\over 3}}}}\over
    {20\,q}} - {{3\,{ A_{1}}\,{ B_{2}}\,
      {t^{{2\over 3}}}}\over {2\,q}} +
  {{763\,{{{ A_{1}}}^2}\,{ A_{2}}\,{t^{{4\over 3}}}}\over
    {72\,{q^2}}}\nonumber\\
&+&\,\left. {{3\,{ C_{3}}\,{t^{{5\over 3}}}}\over
    5} + {{250\,{{{ A_{1}}}^3}\,{K^2}\,
      {t^{{7\over 3}}}}\over {63\,{q^4}}} -
  {{9\,{ A_{1}}\,{ B_{1}}\,{t^{{7\over 3}}}}\over
    {10\,q}} + {{67\,{{{ A_{1}}}^3}\,{t^3}}\over
    {9\,{q^2}}}\right]\,\cos(q\,x)\\
\label{g3,00}
g^{(3)}_{00}&=&\left[{{-5\,{ A_{1}}\,{{{ A_{2}}}^2}\,{K^2}}\over
    {3\,{q^5}\,{t^2}}} -
  {{{ A_{2}}\,{ B_{2}}}\over {2\,{q^2}\,{t^2}}} +
  {{25\,{ A_{1}}\,{{{ A_{2}}}^2}}\over
    {27\,{q^3}\,{t^{{4\over 3}}}}} +
  {{160\,{{{ A_{1}}}^2}\,{ A_{2}}\,{K^2}}\over
    {27\,{q^5}\,{t^{{1\over 3}}}}}\right.\nonumber\\
&+&\,
  {{8\,{ A_{2}}\,{ B_{1}}}\over
    {15\,{q^2}\,{t^{{1\over 3}}}}} +
  {{8\,{ A_{1}}\,{ B_{2}}}\over
    {9\,{q^2}\,{t^{{1\over 3}}}}} -
  {{200\,{{{ A_{1}}}^2}\,{ A_{2}}\,{t^{{1\over 3}}}}\over
    {27\,{q^3}}} + {{2\,{ C_{1}}\,{t^{{2\over 3}}}}\over
    {3\,q}}\nonumber\\
&+&\,\left. {{80\,{{{ A_{1}}}^3}\,{K^2}\,
      {t^{{4\over 3}}}}\over {27\,{q^5}}} +
  {{58\,{ A_{1}}\,{ B_{1}}\,{t^{{4\over 3}}}}\over
    {15\,{q^2}}} - {{350\,{{{ A_{1}}}^3}\,{t^2}}\over
    {27\,{q^3}}}\right]\,\sin(3q\,x)\nonumber\\
&+&\,\left[{{-25\,{ A_{1}}\,{{{ A_{2}}}^2}\,{K^2}}\over
    {9\,{q^5}\,{t^2}}} +
  {{{ A_{2}}\,{ B_{2}}}\over {2\,{q^2}\,{t^2}}} +
  {{{ A_{2}}\,{ F_{1}}}\over {2\,q\,{t^2}}} -
  {{{ A_{2}}\,{ F_{3}}}\over {2\,q\,{t^2}}} -
  {{25\,{ A_{1}}\,{{{ A_{2}}}^2}}\over
    {9\,{q^3}\,{t^{{4\over 3}}}}}\right.\nonumber\\
&-&\,
  {{8\,{ A_{2}}\,{ B_{1}}}\over
    {15\,{q^2}\,{t^{{1\over 3}}}}} -
  {{8\,{ A_{1}}\,{ B_{2}}}\over
    {9\,{q^2}\,{t^{{1\over 3}}}}} -
  {{2\,{ A_{1}}\,{ F_{1}}}\over {q\,{t^{{1\over 3}}}}} +
  {{{ A_{1}}\,{ F_{3}}}\over {3\,q\,{t^{{1\over 3}}}}}\nonumber\\
&+&\,\left.
  {{200\,{{{ A_{1}}}^2}\,{ A_{2}}\,{t^{{1\over 3}}}}\over
    {9\,{q^3}}} + {{2\,{ C_{3}}\,{t^{{2\over 3}}}}\over
    q} + {{1400\,{{{ A_{1}}}^3}\,{K^2}\,
      {t^{{4\over 3}}}}\over {81\,{q^5}}} -
  {{58\,{ A_{1}}\,{ B_{1}}\,{t^{{4\over 3}}}}\over
    {15\,{q^2}}} + {{350\,{{{ A_{1}}}^3}\,{t^2}}\over
    {9\,{q^3}}}\right]\,\sin(q\,x)\\
\label{g3,11}
g^{(3)}_{11}&=&\left[-\left( { A_{2}}\,{ B_{2}} \right)  -
  {{55\,{{{ A_{2}}}^3}\,{K^2}}\over
    {81\,{q^3}\,{t^{{5\over 3}}}}} +
  {{{{{ A_{2}}}^3}}\over {2\,q\,t}} -
  {{1120\,{ A_{1}}\,{{{ A_{2}}}^2}\,{K^4}}\over
    {81\,{q^5}\,{t^{{2\over 3}}}}} -
  {{16\,{ A_{2}}\,{ B_{2}}\,{K^2}}\over
    {9\,{q^2}\,{t^{{2\over 3}}}}}\right.\nonumber\\
&-&\,
  {{1000\,{{{ A_{1}}}^2}\,{ A_{2}}\,{K^6}\,
      {t^{{1\over 3}}}}\over {27\,{q^7}}} -
  {{4\,{ A_{2}}\,{ B_{1}}\,{K^4}\,{t^{{1\over 3}}}}\over
    {9\,{q^4}}} - {{200\,{ A_{1}}\,{ B_{2}}\,{K^4}\,
      {t^{{1\over 3}}}}\over {27\,{q^4}}} +
  {{4\,{ A_{2}}\,{ B_{3}}\,{t^{{1\over 3}}}}\over
    {3\,q}} \nonumber\\
&+&\, {{8\,{ C_{2}}\,{K^2}\,{t^{{1\over 3}}}}\over
      {9\,q}} + {{19\,{ A_{1}}\,{{{ A_{2}}}^2}\,
      {t^{{2\over 3}}}}\over {6\,q}} -
  {{20\,{{{ A_{1}}}^2}\,{ A_{2}}\,{K^4}\,t}\over
    {3\,{q^5}}} - {{2\,{ A_{2}}\,{ B_{1}}\,{K^2}\,t}\over
    {5\,{q^2}}} - {{2\,{ A_{1}}\,{ B_{2}}\,{K^2}\,t}\over
    {3\,{q^2}}}\nonumber\\
&+&\, { C_{6}}\,{t^{{4\over 3}}} -
  {{3\,{ A_{2}}\,{ B_{1}}\,{t^{{5\over 3}}}}\over 5} -
  { A_{1}}\,{ B_{2}}\,{t^{{5\over 3}}} +
  {{5\,{{{ A_{1}}}^2}\,{ A_{2}}\,{K^2}\,
      {t^{{5\over 3}}}}\over {27\,{q^3}}} -
  {{1000\,{{{ A_{1}}}^3}\,{K^6}\,{t^2}}\over
    {27\,{q^7}}} - {{44\,{ A_{1}}\,{ B_{1}}\,{K^4}\,
      {t^2}}\over {9\,{q^4}}}\nonumber\\
&+&\,
  {{4\,{ A_{1}}\,{ B_{3}}\,{t^2}}\over {3\,q}} +
  {{8\,{ C_{1}}\,{K^2}\,{t^2}}\over {15\,q}} +
  {{29\,{{{ A_{1}}}^2}\,{ A_{2}}\,{t^{{7\over 3}}}}\over
    {6\,q}} - {{440\,{{{ A_{1}}}^3}\,{K^4}\,
      {t^{{8\over 3}}}}\over {567\,{q^5}}} +
  {{4\,{ A_{1}}\,{ B_{1}}\,{K^2}\,{t^{{8\over 3}}}}\over
    {15\,{q^2}}} \nonumber\\
&-&\,\left. {{3\,{ A_{1}}\,{ B_{1}}\,
      {t^{{{10}\over 3}}}}\over 5} -
  {{40\,{{{ A_{1}}}^3}\,{K^2}\,{t^{{{10}\over 3}}}}\over
    {81\,{q^3}}} + {{13\,{{{ A_{1}}}^3}\,{t^4}}\over
    {6\,q}}\right]\,\sin(3q\,x)\nonumber\\
&+&\,\left[-\left( { A_{2}}\,{ B_{2}} \right)  +
  {{55\,{{{ A_{2}}}^3}\,{K^2}}\over
    {27\,{q^3}\,{t^{{5\over 3}}}}} +
  {{{{{ A_{2}}}^3}}\over {2\,q\,t}} +
  {{800\,{ A_{1}}\,{{{ A_{2}}}^2}\,{K^4}}\over
    {81\,{q^5}\,{t^{{2\over 3}}}}} +
  {{16\,{ A_{2}}\,{ B_{2}}\,{K^2}}\over
    {9\,{q^2}\,{t^{{2\over 3}}}}}\right.\nonumber\\
&+&\,
  {{4\,{ A_{2}}\,{ F_{1}}\,{K^2}}\over
    {3\,q\,{t^{{2\over 3}}}}} -
  {{2\,{ A_{2}}\,{ F_{3}}\,{K^2}}\over
    {q\,{t^{{2\over 3}}}}} -
  {{25000\,{{{ A_{1}}}^2}\,{ A_{2}}\,{K^6}\,
      {t^{{1\over 3}}}}\over {243\,{q^7}}} +
  {{4\,{ A_{2}}\,{ B_{1}}\,{K^4}\,{t^{{1\over 3}}}}\over
    {3\,{q^4}}}\nonumber\\
&+&\, {{40\,{ A_{1}}\,{ B_{2}}\,{K^4}\,
      {t^{{1\over 3}}}}\over {27\,{q^4}}} +
  {{40\,{ A_{1}}\,{ F_{1}}\,{K^4}\,{t^{{1\over 3}}}}\over
    {9\,{q^3}}} - {{40\,{ A_{1}}\,{ F_{3}}\,{K^4}\,
      {t^{{1\over 3}}}}\over {3\,{q^3}}} -
  {{4\,{ A_{2}}\,{ B_{3}}\,{t^{{1\over 3}}}}\over
    {3\,q}}\nonumber\\
&+&\, {{8\,{ C_{4}}\,{K^2}\,{t^{{1\over 3}}}}\over
      {3\,q}} + {{8\,{ A_{2}}\,{ F_{4}}\,{K^2}\,
      {t^{{1\over 3}}}}\over {3\,q}} +
  {{19\,{ A_{1}}\,{{{ A_{2}}}^2}\,{t^{{2\over 3}}}}\over
    {6\,q}} + {{4540\,{{{ A_{1}}}^2}\,{ A_{2}}\,{K^4}\,
      t}\over {27\,{q^5}}} +
  {{2\,{ A_{2}}\,{ B_{1}}\,{K^2}\,t}\over {5\,{q^2}}}\nonumber\\
&+&\,
  {{2\,{ A_{1}}\,{ B_{2}}\,{K^2}\,t}\over {3\,{q^2}}} + {{18\,{
A_{1}}\,{ F_{1}}\,{K^2}\,t}\over q} +
  {{8\,{ A_{1}}\,{ F_{3}}\,{K^2}\,t}\over q} +
  { C_{5}}\,{t^{{4\over 3}}} -
  {{3\,{ A_{2}}\,{ B_{1}}\,{t^{{5\over 3}}}}\over 5}\nonumber\\
&-&\,
  { A_{1}}\,{ B_{2}}\,{t^{{5\over 3}}} -
  {{5\,{{{ A_{1}}}^2}\,{ A_{2}}\,{K^2}\,
      {t^{{5\over 3}}}}\over {9\,{q^3}}} -
  {{12200\,{{{ A_{1}}}^3}\,{K^6}\,{t^2}}\over
    {243\,{q^7}}} + {{20\,{ A_{1}}\,{ B_{1}}\,{K^4}\,
      {t^2}}\over {9\,{q^4}}}\nonumber\\
&-&
  {{4\,{ A_{1}}\,{ B_{3}}\,{t^2}}\over {3\,q}} +
  {{8\,{ C_{3}}\,{K^2}\,{t^2}}\over {5\,q}} +
  {{8\,{ A_{1}}\,{ F_{4}}\,{K^2}\,{t^2}}\over {3\,q}} +
  {{29\,{{{ A_{1}}}^2}\,{ A_{2}}\,{t^{{7\over 3}}}}\over
    {6\,q}}-{{4360\,{{{ A_{1}}}^3}\,{K^4}\,
      {t^{{8\over 3}}}}\over {567\,{q^5}}}\nonumber\\
&-&\,\left.
  {{4\,{ A_{1}}\,{ B_{1}}\,{K^2}\,{t^{{8\over 3}}}}\over
    {15\,{q^2}}} - {{3\,{ A_{1}}\,{ B_{1}}\,
      {t^{{{10}\over 3}}}}\over 5} +
  {{40\,{{{ A_{1}}}^3}\,{K^2}\,{t^{{{10}\over 3}}}}\over
    {27\,{q^3}}} + {{13\,{{{ A_{1}}}^3}\,{t^4}}\over
    {6\,q}}\right]\,\sin(q\,x)\\
\label{g3,22}
g^{(3)}_{22}&=&\left[{{4\,{ A_{1}}\,{{{ A_{2}}}^2}\,{K^2}}\over
{3\,{q^3}}} +
  {{7\,{{{ A_{2}}}^3}\,{K^2}}\over
    {162\,{q^3}\,{t^{{5\over 3}}}}} -
  {{40\,{ A_{1}}\,{{{ A_{2}}}^2}\,{K^4}}\over
    {81\,{q^5}\,{t^{{2\over 3}}}}} -
  {{4\,{ A_{2}}\,{ B_{2}}\,{K^2}}\over
    {9\,{q^2}\,{t^{{2\over 3}}}}}\right.\nonumber\\
&+&\,
  {{200\,{{{ A_{1}}}^2}\,{ A_{2}}\,{K^6}\,
      {t^{{1\over 3}}}}\over {243\,{q^7}}} -
  {{20\,{ A_{1}}\,{ B_{2}}\,{K^4}\,{t^{{1\over 3}}}}\over
    {27\,{q^4}}} - {{4\,{ C_{2}}\,{K^2}\,
      {t^{{1\over 3}}}}\over {9\,q}} -
  {{110\,{{{ A_{1}}}^2}\,{ A_{2}}\,{K^4}\,t}\over
    {27\,{q^5}}} - {{3\,{ A_{2}}\,{ B_{1}}\,{K^2}\,t}\over
      {5\,{q^2}}} \nonumber\\
&-&\, {{{ A_{1}}\,{ B_{2}}\,{K^2}\,t}\over
    {{q^2}}} - {{68000\,{{{ A_{1}}}^3}\,{K^8}\,
      {t^{{4\over 3}}}}\over {6561\,{q^9}}} +
  {{80\,{ A_{1}}\,{ B_{1}}\,{K^6}\,{t^{{4\over 3}}}}\over
    {243\,{q^6}}} - {{40\,{ A_{1}}\,{ B_{3}}\,{K^2}\,
      {t^{{4\over 3}}}}\over {81\,{q^3}}} -
  {{8\,{ C_{1}}\,{K^4}\,{t^{{4\over 3}}}}\over
    {81\,{q^3}}}\nonumber\\
&+&\, {{187\,{{{ A_{1}}}^2}\,{ A_{2}}\,
      {K^2}\,{t^{{5\over 3}}}}\over {54\,{q^3}}} +
  {{200\,{{{ A_{1}}}^3}\,{K^6}\,{t^2}}\over
    {243\,{q^7}}} - {{4\,{ A_{1}}\,{ B_{1}}\,{K^4}\,
      {t^2}}\over {9\,{q^4}}} -
  {{4\,{ C_{1}}\,{K^2}\,{t^2}}\over {15\,q}} \nonumber\\
&+&\, \left.
  {{220\,{{{ A_{1}}}^3}\,{K^4}\,{t^{{8\over 3}}}}\over
    {567\,{q^5}}} - {{14\,{ A_{1}}\,{ B_{1}}\,{K^2}\,
      {t^{{8\over 3}}}}\over {15\,{q^2}}} +
  {{176\,{{{ A_{1}}}^3}\,{K^2}\,{t^{{{10}\over 3}}}}\over
    {81\,{q^3}}}\right]\,\sin(3q\,x)\nonumber\\
&+&\,\left[{{-4\,{ A_{1}}\,{{{ A_{2}}}^2}\,{K^2}}\over {{q^3}}} -
  {{7\,{{{ A_{2}}}^3}\,{K^2}}\over
    {54\,{q^3}\,{t^{{5\over 3}}}}} -
  {{280\,{ A_{1}}\,{{{ A_{2}}}^2}\,{K^4}}\over
    {81\,{q^5}\,{t^{{2\over 3}}}}} +
  {{4\,{ A_{2}}\,{ B_{2}}\,{K^2}}\over
    {9\,{q^2}\,{t^{{2\over 3}}}}}\right.\nonumber\\
&+&\,
  {{2\,{ A_{2}}\,{ F_{1}}\,{K^2}}\over
    {3\,q\,{t^{{2\over 3}}}}} -
  {{{ A_{2}}\,{ F_{3}}\,{K^2}}\over
    {3\,q\,{t^{{2\over 3}}}}} +
  {{800\,{{{ A_{1}}}^2}\,{ A_{2}}\,{K^6}\,
      {t^{{1\over 3}}}}\over {27\,{q^7}}} -
  {{4\,{ A_{2}}\,{ B_{1}}\,{K^4}\,{t^{{1\over 3}}}}\over
    {9\,{q^4}}} + {{100\,{ A_{1}}\,{ B_{2}}\,{K^4}\,
      {t^{{1\over 3}}}}\over {27\,{q^4}}} \nonumber\\
&+&\,
  {{20\,{ A_{1}}\,{ F_{1}}\,{K^4}\,{t^{{1\over 3}}}}\over
    {9\,{q^3}}} + {{20\,{ A_{1}}\,{ F_{3}}\,{K^4}\,
      {t^{{1\over 3}}}}\over {9\,{q^3}}} -
  {{4\,{ C_{4}}\,{K^2}\,{t^{{1\over 3}}}}\over {3\,q}} -
  {{4\,{ A_{2}}\,{ F_{2}}\,{K^2}\,{t^{{1\over 3}}}}\over
    {3\,q}} - {{130\,{{{ A_{1}}}^2}\,{ A_{2}}\,{K^4}\,
      t}\over {9\,{q^5}}} \nonumber\\
&+&\,
  {{3\,{ A_{2}}\,{ B_{1}}\,{K^2}\,t}\over {5\,{q^2}}} +
  {{{ A_{1}}\,{ B_{2}}\,{K^2}\,t}\over {{q^2}}} +
  {{7\,{ A_{1}}\,{ F_{1}}\,{K^2}\,t}\over {3\,q}} -
  {{{ A_{1}}\,{ F_{3}}\,{K^2}\,t}\over {3\,q}} +
  {{82000\,{{{ A_{1}}}^3}\,{K^8}\,{t^{{4\over 3}}}}\over
    {729\,{q^9}}}\nonumber\\
&-&\, {{40\,{ A_{1}}\,{ B_{1}}\,{K^6}\,
      {t^{{4\over 3}}}}\over {27\,{q^6}}} -
  {{8\,{ C_{3}}\,{K^4}\,{t^{{4\over 3}}}}\over
    {3\,{q^3}}} - {{40\,{ A_{1}}\,{ F_{2}}\,{K^4}\,
      {t^{{4\over 3}}}}\over {9\,{q^3}}}\nonumber\\
&-&\,
  {{40\,{ A_{1}}\,{ F_{4}}\,{K^4}\,{t^{{4\over 3}}}}\over
    {9\,{q^3}}} - {{187\,{{{ A_{1}}}^2}\,{ A_{2}}\,{K^2}\,
      {t^{{5\over 3}}}}\over {18\,{q^3}}} +
  {{800\,{{{ A_{1}}}^3}\,{K^6}\,{t^2}}\over
    {243\,{q^7}}} + {{16\,{ A_{1}}\,{ B_{1}}\,{K^4}\,
      {t^2}}\over {9\,{q^4}}}\nonumber\\
&-&\,
  {{4\,{ C_{3}}\,{K^2}\,{t^2}}\over {5\,q}} -
  {{4\,{ A_{1}}\,{ F_{2}}\,{K^2}\,{t^2}}\over {3\,q}} -

  {{9580\,{{{ A_{1}}}^3}\,{K^4}\,{t^{{8\over 3}}}}\over
    {567\,{q^5}}} + {{14\,{ A_{1}}\,{ B_{1}}\,{K^2}\,
      {t^{{8\over 3}}}}\over {15\,{q^2}}}\nonumber\\
&-&\,\left.
  {{176\,{{{ A_{1}}}^3}\,{K^2}\,{t^{{{10}\over 3}}}}\over
    {27\,{q^3}}}\right]\,\sin\,(q\,x)
\end{eqnarray}
The new parameters ( or integration constants) $C_{i}$ arise when we
solve the homogeneous part of the differential equation system
(\ref{eqs,n}). Terms, which contain the already known
$A_{i}$,$B_{i}$, or $F_{i}$, are generated as special inhomogeneous
solutions of (\ref{eqs,n}) by the source terms $S_{\mu\nu}^{(2)}$.
Note that in third order the wave number has tripled compared with
the first order; additionally, terms proportional to $\sin(qx)$ or
$\cos(qx)$ arise.\\

The following, general structure of the PMF solution for the dust
universe now becomes visible:
\begin{eqnarray}
\label{allgemeinestaubloesung}
g_{\mu\nu}^{(n)} &=& f_{0}^{(n)}[t]trig(nqx) +
f_{2}^{(n)}[t]trig((n-2)qx) + f_{4}^{(n)}[t]trig((n-4)qx) + \cdots
+\nonumber\\
& & f_{2Int[n/2]}^{(n)}[t]trig((n-2Int[n/2])qx).\nonumber\\
\end{eqnarray}
Here, the $f_{i}^{(n)}[t]$ are sums of powers of $t$; $trig$ is
either $\sin$ or $\cos$ ( according to the order and the component in
question), and $Int$ is the integer function. Thus, in $n^{th}$ order
we get the wave numbers nq, (n-2)q, (n-4)q,...; this sequence is
ending with $q$ or $0q$. Hence, a harmonic (i.e., proportional to
$\sin$ or $\cos$) fluctuation is, in higher orders, necessarily
accompanied by corrections of equal and lesser extension. It must be
mentioned that (\ref{allgemeinestaubloesung}) still has to be proved
for general $n$. Comparison of the functions $f_{i}^{(n)}[t]$ for n =
1,2,3 with each other shows regularities which we are going to
discuss now by means of the following case study.\\

To that end we set the free parameters of orders higher than the
first one equal to zero; hence, the only remaining non-vanishing
parameters are those given by the first order, namely the $A_{i}$'s.
Let us write down here merely the 10-component:\\
\begin{eqnarray}
\label{g1,10,fs}
g^{(1)}_{10}&=&\left( { A_{2}} + { A_{1}}\,{t^{{5\over 3}}} \right)
\,
  \cos (q\,x)\\
\label{g2,10,fs}
g^{(2)}_{10}&=&\left[ - {{{{{ A_{2}}}^2}}\over {2\,q\,t}} -
  {{7\,{ A_{1}}\,{ A_{2}}\,{t^{{2\over 3}}}}\over
    {2\,q}}-
  {{3\,{{{ A_{1}}}^2}\,{t^{{7\over 3}}}}\over
q}\right]\,\sin(2q\,x)\\
\label{g3,10,fs}
g^{(3)}_{10}&=&\left[{{{{{ A_{2}}}^3}}\over {8\,{q^2}\,{t^2}}} -
{{109\,{ A_{1}}\,{{{ A_{2}}}^2}}\over
    {36\,{q^2}\,{t^{{1\over 3}}}}}- {{763\,{{{ A_{1}}}^2}\,{
A_{2}}\,{t^{{4\over 3}}}}\over
    {72\,{q^2}}} - {{67\,{{{ A_{1}}}^3}\,{t^3}}\over
    {9\,{q^2}}} \right.\nonumber\\
    &+&\, \left.
     {{5\,{ A_{1}}\,{{{ A_{2}}}^2}\,{K^2}}\over
    {2\,{q^4}\,t}}+{{40\,{{{ A_{1}}}^2}\,{ A_{2}}\,{K^2}\,
      {t^{{2\over 3}}}}\over {3\,{q^4}}} + {{40\,{{{
A_{1}}}^3}\,{K^2}\,
      {t^{{7\over 3}}}}\over {21\,{q^4}}}
\right]\,\cos(3q\,x)\nonumber\\
&+&\,\left[ - {{{{{ A_{2}}}^3}}\over {8\,{q^2}\,{t^2}}}  + {{109\,{
A_{1}}\,{{{ A_{2}}}^2}}\over
    {36\,{q^2}\,{t^{{1\over 3}}}}}+
  {{763\,{{{ A_{1}}}^2}\,{ A_{2}}\,{t^{{4\over 3}}}}\over
    {72\,{q^2}}}+ {{67\,{{{ A_{1}}}^3}\,{t^3}}\over
    {9\,{q^2}}}\right.\nonumber\\
 &+&\, \left.
  {{25\,{ A_{1}}\,{{{ A_{2}}}^2}\,{K^2}}\over
    {18\,{q^4}\,t}}+{{25\,{{{ A_{1}}}^2}\,{ A_{2}}\,{K^2}\,
      {t^{{2\over 3}}}}\over {27\,{q^4}}}+{{250\,{{{
A_{1}}}^3}\,{K^2}\,
      {t^{{7\over 3}}}}\over {63\,{q^4}}}  \right]\,\cos(q\,x)
\end{eqnarray}
First of all, the results (\ref{g1,10,fs}), (\ref{g2,10,fs}), and
(\ref{g3,10,fs}) show again very clearly that in higher orders {\it
necessarily} corrections to the first order arise (observe, that the
$g_{10}^{(n)}$ are different from zero for $n \geq 1$).\\
We have written down the 10-component in each order such that within
that order the sums are ordered according to different powers of
$1/q$; and within these powers they are ordered according to
increasing powers of $A_{1}$. Now we see the following: within the
same power of $1/q$ the exponent of $t$ is increasing by $5/3$ when
the power of $A_{1}$ increases by 1. From one order to the next the
lowest power of $1/q$ which appears increases always by $1/q$, the
corresponding sequence of exponents of $t$ is starting from a value
which is reduced by 1 ( compared with the starting value of the
previous order): $A_{2}t^{0}$, ${A_{2}}^{2}t^{-1}$,
${A_{2}}^{3}t^{-2}$,etc. The highest time exponent appearing in each
order increases order by order by $2/3$: $A_{1}t^{\frac{5}{3}}$,
${A_{1}}^{2}t^{\frac{7}{3}}$, ${A_{1}}^{3}t^{\frac{9}{3}}$. In the
third order we have additionally to the terms proportional to $1/q^2$
such proportional to $K^2/q^4$. They have the same time exponents as
the terms of the second order where the one with the highest time
exponent (proportional to ${A_{1}}^{3}K^2t^{\frac{7}{3}}/q^4$)
corresponds to that one with the highest time exponent in second
order (proportional to ${A_{1}}^{2}t^{\frac{7}{3}}/q$). Since in
second order there are only three such terms available, we have no
contribution proportional to ${A_{2}}^{3}K^2/q^4$. This holds as well
for the terms proportional to $\cos(3q\,x)$ as for those proportional
to $\cos(q\,x)$. Thus, we can expect the following terms in fourth
order:\\

proportional to $1/q^3$:\\
${A_{2}}^{4}t^{-3}/q^{3}$,
${A_{2}}^{3}{A_{1}}t^{\frac{-4}{3}}/q^{3}$,
${A_{2}}^{2}{A_{1}}^{2}t^{\frac{1}{3}}/q^{3}$,
${A_{2}}{A_{1}}^{3}t^{2}/q^{3}$,
${A_{1}}^{4}t^{\frac{11}{3}}/q^{3}$,\\

proportional to $K^2/q^5$:\\
${A_{2}}^{3}{A_{1}}t^{-2}K^{2}/q^{5}$,
${A_{2}}^{2}{A_{1}}^{2}t^{\frac{-1}{3}}K^{2}/q^{5}$,
${A_{2}}{A_{1}}^{3}t^{\frac{4}{3}}K^{2}/q^{5}$,
${A_{1}}^{4}t^{3}K^{2}/q^{5}$,\\

proportional to $K^4/q^7$:\\
${A_{2}}^{2}{A_{1}}^{2}t^{-1}K^{4}/q^{7}$,
${A_{2}}{A_{1}}^{3}t^{\frac{2}{3}}K^{4}/q^{7}$,
${A_{1}}^{4}t^{\frac{7}{3}}K^{4}/q^{7}$.\\

(the appearance of the terms proportional to $K^4/q^7$ is especially
suggested by doing an analogeous consideration for the other metric
components).\\
This consideration shows that fluctuations with large spatial
extension (i.e., $q\rightarrow 0$) are governed within each order $n$
by terms ${A_{1}}^{n}K^{2(n-2)}t^{\frac{7}{3}}/q^{3n-5}$, if we
consider here only the growing mode (to this end we set $A_{2}=0$),
for it possesses the highest power of $1/q$. For smaller fluctuations
also terms with smaller powers of $1/q$ are important. But those
contain higher powers of $t$! The one with the highest power of $t$
is proportional to ${A_{1}}^{n}t^{\frac{2n}{3}+1}/q^{n-1}$. This
means that such fluctuations are growing faster provided that we are
far beyond the Jeans limit (this lower boundary of the instability
region is in the case of the dust universe simply $q=\infty$). Let us
stress that while in first order small extended fluctuations grow as
fast as large extented ones (namely, proportional to
$A_{1}t^{\frac{5}{3}}$), a higher-order analysis shows that they have
different growth rates. The reason is that order by order terms with
constantly increasing time exponents are appearing; but these "growth
terms" are dominated in the case of perturbations with large
extension by others which are growing only moderately. It is true
that this dominance is dissappearing when $t$ becomes sufficiently
big enough but this boundary can be shifted arbitrarily towards the
future if we merely take $q$ small enough. This is a clear hint for
the existence of an upper boundary of the region of instability, and
this existence is caused by non-linear effects. To sum up: it seems
that non-linear effects do not stabilize a perturbation which is
unstable according to the first-order analysis but they let large
extented fluctuations just moderately grow while they cause small
extented ones to grow much faster.\\

The other metric components show an analogeous behaviour. It must be
mentioned that all those conjectures have not been proven yet
strictly; they  have just been  verified for $n \leq 3$. This should
be done in the future; similary, we have to find the algorithm for
the numerical coefficients in our sums. Especially for small extented
fluctuation the whole sum is important and we have to calculate its
limit in order to know what functions of $t$ the metric components
are. However, since those "growth terms", which are proportinal to
$\cos(qx)$, namely $A_{1}t^{\frac{5}{3}}$ in $g_{10}^{(1)}$ and
${{67\,{{{ A_{1}}}^3}\,{t^3}}\over
    {9\,{q^2}}}$ in $g_{10}^{(3)}$, both have the same sign "+", one
might expect that summing up all those contributions we get an
rapidly increasing perturbation (provided, of course, that q is not
too small). Moreover, our analysis has been performed so far merely
within the PMF gauge. It remains to study whether the results found
in that gauge, in particular to what extend the non-linear effects
survive if we transform the PMF solution into a gauge which is "close
to the background"\\

One can also infer from (\ref{g1,10,fs}), (\ref{g2,10,fs}), and
(\ref{g3,10,fs}) the following feature. Let us assume that the
first-order perturbation quantities $g_{\mu\nu}^{(1)}$ are small
compared to the background quantities (i.e., the parameters $A_{i}$
have to be very small). Then, the necessary correction terms of
higher orders like $g_{\mu\nu}^{(2)}$, $g_{\mu\nu}^{(3)}$, etc. are
automatically small compared to those of lower orders provided that
$q$ and $t$ are moderate. The reason is that they contain terms
proportional to ${A_{1}}^r{A_{2}}^s$,  where $r+s=n$. Hence, for
small times and for not too largely extented fluctuations a first
order analysis is justified. This is what we expect.

\section{Conclusions and Perspectives}
Let us now summarize the main results of our higher-order analysis.
First of all higher order ("non-linear") effects cause, in principle,
perturbations to grow much faster than they grow according to the
first-order analysis. The reason for that behaviour is that higher
order contributions contain "growth terms", and it seems that they
all have the same sign (i.e., they are all acting along the same
direction). However, for very large perturbations those growth terms
are dominated within each order by others which grow only moderately.
Thus, we get the following picture: fluctuations with large extension
(beyond super-clusters of galaxies?) grow only moderately (more or
less with a similar  small rate they grow according to the
first-order analysis) but the other perturbations grow much faster
provided their extension is beyond the Jeans limit. It must be
mentioned that this interesting result has to be taken for the moment
just as a conjecture which is supported by some tendencies observed
from the results in PMF gauge in first, second, and third order. We
still have to work out the solution in $n^{th}$ order. Moreover, our
equation of state used here (dust universe) is not too realistic with
regard to the formation of galaxies. And, finally, in order to be in
the position for judging stability/instability of a given
perturbation we should transform our PMF solution into some
appropriate gauge. Then, we can infer from the density contrast
within this gauge whether that perturbation is stable or not.
However, since in the PMF gauge instabilities show up in the metric
components alone rather than in the density contrast, one can expect
that in an appropriate gauge we will observe a similar behaviour for
the density contrast like that which the metric components show in
the PMF gauge. If, after performing all these improvements and
transformations, the observed tendencies survive at least in
principle, we have found a possible explanation for the breaking off
of the hierachie [clusters of stars - galaxies - clusters of galaxies
- super-clusters] at  super-clusters or at super-super-clusters
\cite{astronomieI}, \cite{astronomieII}, \cite{astronomieIII},
\cite{gottI}, \cite{gottII}. In this case, general relativity would
imply an upper boundary of the instability region.\\

Another result is that the transition to higher orders is connected
with a multiplication of the wave numbers. Maximally, we obtain in
$n^{th}$ order a contribution proportional to $\sin(nqx)$ or
$\cos(nqx)$; but there are also terms with smaller wave numbers (see
(\ref{allgemeinestaubloesung})). Altogether, we observe a kind of
fragmantation which is increasing with the order (i.e., with the
evolution time). That means, that more and more parts of the
fluctuation evolve differently, and the extension of those parts
become smaller and smaller when the order increases. Hence, the
perturbation is fraying more and more when time passes by.\\

In our analysis we have considered only fluctuations with a spatial
shape proportional to sine or cosine. However, the realistic
fluctuations are those whose spatial shape is e. g. something like a
Gaussian. This is not at all a problem in first order  because we can
obtain the solution for any shape by means of a Fourier synthesis.
But in the non-linear theory, a sum of solutions of the field
equations is not necessarily also a solution of these equations.
Nevertheless, we can use our solutions obtained in this paper also in
such a case. We just have to perform the Fourier analysis "order by
order". This expression is to be understood as follows. First of all,
we perform the Fourier analysis in first order. Our solutions
proportional to $exp(iqx)$ are solutions of the individual Fourier
components. Subsequently, we compose them and obtain the solution of
the given perturbation in first order. That solution must then be
inserted into the source terms in (\ref{eqs,n}) for $n = 2$. We form
the products and sums of $S_{\mu\nu}^{(1)}$, and, after that, we
decompose the source terms into their Fourier components. Then, our
solutions proportional to $exp(iqx)$ are again solutions of the
individual Fourier components. Since the perturbation field equations
(\ref{eqs,n}) are also in second order linear in the unknown
functions $g_{\mu\nu}^{(2)}$, the Fourier synthesis out of our
solutions proportional to sine and cosine yields the full solution in
second order which is to be inserted into the source terms in third
order. In this way we proceed order by order.\\
Moreover, one should rid oneself of the concept of perturbations with
two-dimensional symmetry planes,  if the aim is a theory of the
formation of galaxies as realistic as possible. Instead, one should
consider perturbations which are (approximately) spherically
symmetric. The transition to these fluctuations should not be any
problem because the starting metric (in spherical coordinates) is not
much more complicated than (\ref{robertsonwalker}). Our PMF methods
should, hence, be applicable also in this case.\\

The PMF method can be used also on a larger scale. If our conjecture
about the solution in an arbitrary order turns out to be true, we can
generate by means of the PMF method exact solutions (presentated as
infinite power series in $t$) of Einstein's field equations. Then, it
is not necessary to assume that the fluctuations are small compared
to the corresponding background quantities; however, if we drop that
assumption, all orders are equally important. An investigation about
the structure of these solutions and their classification could give
useful information about classical General Relativity.\\

A further application is obvious. If we give up our separation ansatz
$g_{\mu\nu}^{(n)} = f(t)h(x)$, we can investigate also gravitational
waves propagating in a Friedman-Robertson-Walker universe, i.e.,
propagation through matter! It should be possible to obtain solutions
(with the help of the PMF method) even without such an separation
ansatz, because our system of differential equations (\ref{eqs,n})
can be decoupled independently from that ansatz (see e.g.
\cite{paperI},(3.17-3.22)).\\

Finally, we want to emphasize that the main aim of this paper is not
to supply a theory of galaxy formation, which is as realistic as
possible. These investigations should rather be understood as a first
step towards such a theory satisfying astrophysicists. First of all,
we are interested in the development of a method which is powerful
enough for solving the field equations also in the case of
space-times, which are less homogeneous than, e.g., the Friedman
universe. We wanted to understand what kind of principle problems
arise, and how they can be handled. Additionally, we were interested
in what kind of new effects caused by higher orders appear. The most
essential new non-linear effect is that perturbations grow much
faster than they do according to a first-order analysis, but those
perturbations with an extremely large extension do not.\\

{\bf Acknowledgements}\\
The author wishes to express his thanks to Erland Wittk\"otter for
his advices concerning the application of some mathematical software
programs which helped to do the necessary calculations in much
shorter time. Additionally, he wants to thank Heinz Dehnen for some
discussions, Stephan Hartmann for reading the manuscript,  and the
Deutsche Forschungsgemeinschaft for financial support.\\
\begin{appendix}
\section{Appendix}
Here we give the full perturbation field equations up to second order
for the relevant components:\\

{\bf First Order:}\\
\\
00 component:\\
\begin{eqnarray}
\label{eq00,1}
&&8\,{ G}\,\pi \,{ g^{(1)}_{00}}(x,t)\,{ \rho_{0}}(t) +
  {{4\,{ g^{(1)}_{22}}(x,t)\,{{\dot{R}(t)}^2}}\over {{{R(t)}^4}}} +
  {{2\,{ g^{(1)}_{11}}(x,t)\,{{\dot{R}(t)}^2}}\over {{{R(t)}^4}}} -
  {{6\,{ g^{(1)}_{00}}(x,t)\,{{\dot{R}(t)}^2}}\over
{{{R(t)}^2}}}\nonumber\\
&&- {{2\,\dot{R}(t)\,\frac{dg^{(1)}_{22}(x,t)}{dt}}\over
{{{R(t)}^3}}} -
  {{\dot{R}(t)\,\frac{dg^{(1)}_{11}(x,t)}{dt}}\over {{{R(t)}^3}}} +
  {{2\,\dot{R}(t)\,\frac{dg^{(1)}_{10}(x,t)}{dx}}\over
{{{R(t)}^3}}}\nonumber\\
 &&+ {{\frac{d^{2}g^{(1)}_{22}(x,t)}{dx^{2}}}\over {{{R(t)}^4}}} = 0
\end{eqnarray}
10 component:\\
\begin{eqnarray}
\label{eq10,1}
&&{{8\,{ G}\,\pi \,{ g^{(1)}_{10}}(x,t)\,{ p_{0}}(t)}\over
    {{{R(t)}^2}}} + {{3\,{ g^{(1)}_{10}}(x,t)\,{{\dot{R}(t)}^2}}\over
    {{{R(t)}^4}}} - {{\dot{R}(t)\,\frac{dg^{(1)}_{00}(x,t)}{dx}}\over
    {{{R(t)}^3}}} + {{2\,\dot{R}(t)\,
      \frac{dg^{(1)}_{22}(x,t)}{dx}}\over {{{R(t)}^5}}} -
  {{\frac{d^{2}g^{(1)}_{22}(x,t)}{dxdt}}\over
{{{R(t)}^4}}}\nonumber\\
&&= 0
\end{eqnarray}
energy conservation:\\
\begin{equation}
\label{ec,1}
{{-4\,{ g^{(1)}_{22}}(x,t)\,\dot{R}(t)}\over {{{R(t)}^3}}} -
  {{2\,{ g^{(1)}_{11}}(x,t)\,\dot{R}(t)}\over {{{R(t)}^3}}} +
  {{2\,\frac{dg^{(1)}_{22}(x,t)}{dt}}\over {{{R(t)}^2}}} +
  {{\frac{dg^{(1)}_{11}(x,t)}{dt}}\over {{{R(t)}^2}}} = 0
\end{equation}
momentum conservation:\\
\begin{equation}
\label{mc,1}
{{{ g^{(1)}_{10}}(x,t)\,\dot{p}_{0}(t)}\over {{{R(t)}^2}}} +
  {{{ p_{0}}(t)\,\frac{dg^{(1)}_{10}(x,t)}{dt}}\over
    {{{R(t)}^2}}} + {{{ \rho_{0}}(t)\,
      \frac{dg^{(1)}_{10}(x,t)}{dt}}\over {{{R(t)}^2}}} -
  {{{ p_{0}}(t)\,\frac{dg^{(1)}_{00}(x,t)}{dx}}\over
    {2\,{{R(t)}^2}}} - {{{ \rho_{0}}(t)\,
      \frac{dg^{(1)}_{00}(x,t)}{dx}}\over {2\,{{R(t)}^2}}} = 0
\end{equation}
{\bf Second Order:}\\
\\
00 component:\\
\begin{eqnarray}
\label{eq00,2}
&&8\,{ G}\,\pi \,{ g^{(2)}_{00}}(x,t)\,{ \rho_{0}}(t) +
  {{4\,{ g^{(2)}_{22}}(x,t)\,{{\dot{R}(t)}^2}}\over {{{R(t)}^4}}} +
  {{2\,{ g^{(2)}_{11}}(x,t)\,{{\dot{R}(t)}^2}}\over {{{R(t)}^4}}} -
  {{6\,{ g^{(2)}_{00}}(x,t)\,{{\dot{R}(t)}^2}}\over
{{{R(t)}^2}}}\nonumber\\
&& - {{2\,\dot{R}(t)\,\frac{dg^{(2)}_{22}(x,t)}{dt}}\over
{{{R(t)}^3}}} -
  {{\dot{R}(t)\,\frac{dg^{(2)}_{11}(x,t)}{dt}}\over {{{R(t)}^3}}} +
  {{2\,\dot{R}(t)\,\frac{dg^{(2)}_{10}(x,t)}{dx}}\over {{{R(t)}^3}}}
+
  {{\frac{d^{2}g^{(2)}_{22}(x,t)}{dx^{2}}}\over
{{{R(t)}^4}}}\nonumber\\
&& + {{8\,{ G}\,\pi \,{{{ g^{(1)}_{10}}(x,t)}^2}\,
      { p_{0}}(t)}\over {{{R(t)}^2}}} +
  8\,{ G}\,\pi \,{{{ g^{(1)}_{00}}(x,t)}^2}\,
   { \rho_{0}}(t) - {{5\,{{{ g^{(1)}_{22}}(x,t)}^2}\,
      {{\dot{R}(t)}^2}}\over {{{R(t)}^6}}} \nonumber\\
&&-
  {{2\,{ g^{(1)}_{22}}(x,t)\,{
g^{(1)}_{11}}(x,t)\,{{\dot{R}(t)}^2}}\over
    {{{R(t)}^6}}}- {{2\,{{{ g^{(1)}_{11}}(x,t)}^2}\,
      {{\dot{R}(t)}^2}}\over {{{R(t)}^6}}} +
  {{6\,{{{ g^{(1)}_{10}}(x,t)}^2}\,{{\dot{R}(t)}^2}}\over
    {{{R(t)}^4}}}\nonumber\\
&& + {{8\,{ g^{(1)}_{00}}(x,t)\,{ g^{(1)}_{22}}(x,t)\,
      {{\dot{R}(t)}^2}}\over {{{R(t)}^4}}}  +
  {{4\,{ g^{(1)}_{00}}(x,t)\,{
g^{(1)}_{11}}(x,t)\,{{\dot{R}(t)}^2}}\over
    {{{R(t)}^4}}}\nonumber\\
&& - {{9\,{{{ g^{(1)}_{00}}(x,t)}^2}\,
      {{\dot{R}(t)}^2}}\over {{{R(t)}^2}}}  +
  {{3\,{
g^{(1)}_{22}}(x,t)\,\dot{R}(t)\,\frac{dg^{(1)}_{22}(x,t)}{dt}}\over
    {{{R(t)}^5}}} + {{{ g^{(1)}_{11}}(x,t)\,\dot{R}(t)\,
      \frac{dg^{(1)}_{22}(x,t)}{dt}}\over {{{R(t)}^5}}}\nonumber\\
&& -
  {{4\,{
g^{(1)}_{00}}(x,t)\,\dot{R}(t)\,\frac{dg^{(1)}_{22}(x,t)}{dt}}\over
    {{{R(t)}^3}}} - {{{{\frac{dg^{(1)}_{22}(x,t)}{dt}}^2}}\over
    {4\,{{R(t)}^4}}}+
  {{{
g^{(1)}_{22}}(x,t)\,\dot{R}(t)\,\frac{dg^{(1)}_{11}(x,t)}{dt}}\over
    {{{R(t)}^5}}} +\nonumber\\
&& {{{ g^{(1)}_{11}}(x,t)\,\dot{R}(t)\,
      \frac{dg^{(1)}_{11}(x,t)}{dt}}\over {{{R(t)}^5}}} -
  {{2\,{
g^{(1)}_{00}}(x,t)\,\dot{R}(t)\,\frac{dg^{(1)}_{11}(x,t)}{dt}}\over
    {{{R(t)}^3}}} - {{\frac{dg^{(1)}_{22}(x,t)}{dt}\,
      \frac{dg^{(1)}_{11}(x,t)}{dt}}\over {2\,{{R(t)}^4}}}
\nonumber\\
&& -
  {{2\,{
g^{(1)}_{22}}(x,t)\,\dot{R}(t)\,\frac{dg^{(1)}_{10}(x,t)}{dx}}\over
    {{{R(t)}^5}}} - {{2\,{ g^{(1)}_{11}}(x,t)\,\dot{R}(t)\,
      \frac{dg^{(1)}_{10}(x,t)}{dx}}\over {{{R(t)}^5}}} +
  {{4\,{
g^{(1)}_{00}}(x,t)\,\dot{R}(t)\,\frac{dg^{(1)}_{10}(x,t)}{dx}}\over
    {{{R(t)}^3}}} \nonumber\\
&& + {{\frac{dg^{(1)}_{22}(x,t)}{dt}\,
      \frac{dg^{(1)}_{10}(x,t)}{dx}}\over {{{R(t)}^4}}} + {{2\,{
g^{(1)}_{10}}(x,t)\,\dot{R}(t)\,
      \frac{dg^{(1)}_{22}(x,t)}{dx}}\over {{{R(t)}^5}}} -
  {{{{\frac{dg^{(1)}_{22}(x,t)}{dx}}^2}}\over {4\,{{R(t)}^6}}}
\nonumber\\
&&-
  {{{
g^{(1)}_{10}}(x,t)\,\dot{R}(t)\,\frac{dg^{(1)}_{11}(x,t)}{dx}}\over
    {{{R(t)}^5}}} - {{\frac{dg^{(1)}_{22}(x,t)}{dx}\,
      \frac{dg^{(1)}_{11}(x,t)}{dx}}\over {2\,{{R(t)}^6}}} -
  {{{ g^{(1)}_{22}}(x,t)\,\frac{d^{2}g^{(1)}_{22}(x,t)}{dx^{2}}}\over
    {{{R(t)}^6}}}\nonumber\\
&& - {{{ g^{(1)}_{11}}(x,t)\,
      \frac{d^{2}g^{(1)}_{22}(x,t)}{dx^{2}}}\over {{{R(t)}^6}}} +
  {{{ g^{(1)}_{00}}(x,t)\,\frac{d^{2}g^{(1)}_{22}(x,t)}{dx^{2}}}\over
    {{{R(t)}^4}}} = 0
\end{eqnarray}
10 component:\\
\begin{eqnarray}
\label{eq10,2}
&&{{8\,{ G}\,\pi \,{ g^{(2)}_{10}}(x,t)\,{ p_{0}}(t)}\over
    {{{R(t)}^2}}} + {{3\,{ g^{(2)}_{10}}(x,t)\,{{\dot{R}(t)}^2}}\over
    {{{R(t)}^4}}} - {{\dot{R}(t)\,\frac{dg^{(2)}_{00}(x,t)}{dx}}\over
    {{{R(t)}^3}}} \nonumber\\
&&+ {{2\,\dot{R}(t)\,
      \frac{dg^{(2)}_{22}(x,t)}{dx}}\over {{{R(t)}^5}}} -
  {{\frac{d^{2}g^{(2)}_{22}(x,t)}{dxdt}}\over {{{R(t)}^4}}} -
{{8\,{ G}\,\pi \,{ g^{(1)}_{10}}(x,t)\,{ g^{(1)}_{11}}(x,t)\,
      { p_{0}}(t)}\over {{{R(t)}^4}}}\nonumber\\
 &&+
  {{8\,{ G}\,\pi \,{ g^{(1)}_{00}}(x,t)\,{ g^{(1)}_{10}}(x,t)\,
      { p_{0}}(t)}\over {{{R(t)}^2}}} - {{4\,{ g^{(1)}_{10}}(x,t)\,{
g^{(1)}_{22}}(x,t)\,
      {{\dot{R}(t)}^2}}\over {{{R(t)}^6}}} -
  {{5\,{ g^{(1)}_{10}}(x,t)\,{
g^{(1)}_{11}}(x,t)\,{{\dot{R}(t)}^2}}\over
    {{{R(t)}^6}}} \nonumber\\
&&+ {{6\,{ g^{(1)}_{00}}(x,t)\,{ g^{(1)}_{10}}(x,t)\,
      {{\dot{R}(t)}^2}}\over {{{R(t)}^4}}}+
  {{2\,{
g^{(1)}_{10}}(x,t)\,\dot{R}(t)\,\frac{dg^{(1)}_{22}(x,t)}{dt}}\over
    {{{R(t)}^5}}} + {{{ g^{(1)}_{10}}(x,t)\,\dot{R}(t)\,
      \frac{dg^{(1)}_{11}(x,t)}{dt}}\over {{{R(t)}^5}}} \nonumber\\
&&+
  {{{
g^{(1)}_{22}}(x,t)\,\dot{R}(t)\,\frac{dg^{(1)}_{00}(x,t)}{dx}}\over
    {{{R(t)}^5}}} + {{{ g^{(1)}_{11}}(x,t)\,\dot{R}(t)\,
      \frac{dg^{(1)}_{00}(x,t)}{dx}}\over {{{R(t)}^5}}} -
  {{2\,{
g^{(1)}_{00}}(x,t)\,\dot{R}(t)\,\frac{dg^{(1)}_{00}(x,t)}{dx}}\over
    {{{R(t)}^3}}}\nonumber\\
&&-
{{\frac{dg^{(1)}_{22}(x,t)}{dt}\,
      \frac{dg^{(1)}_{00}(x,t)}{dx}}\over {2\,{{R(t)}^4}}} -
  {{3\,{
g^{(1)}_{22}}(x,t)\,\dot{R}(t)\,\frac{dg^{(1)}_{22}(x,t)}{dx}}\over
    {{{R(t)}^7}}} - {{3\,{ g^{(1)}_{11}}(x,t)\,\dot{R}(t)\,
      \frac{dg^{(1)}_{22}(x,t)}{dx}}\over {{{R(t)}^7}}}\nonumber\\
&& +
  {{2\,{
g^{(1)}_{00}}(x,t)\,\dot{R}(t)\,\frac{dg^{(1)}_{22}(x,t)}{dx}}\over
    {{{R(t)}^5}}} + {{\frac{dg^{(1)}_{22}(x,t)}{dt}\,
      \frac{dg^{(1)}_{22}(x,t)}{dx}}\over {2\,{{R(t)}^6}}} +

{{\frac{dg^{(1)}_{11}(x,t)}{dt}\,\frac{dg^{(1)}_{22}(x,t)}{dx}}\over
    {2\,{{R(t)}^6}}}\nonumber\\
&&+
  {{{ g^{(1)}_{22}}(x,t)\,\frac{d^{2}g^{(1)}_{22}(x,t)}{dxdt}}\over
    {{{R(t)}^6}}} + {{{ g^{(1)}_{11}}(x,t)\,
      \frac{d^{2}g^{(1)}_{22}(x,t)}{dxdt}}\over {{{R(t)}^6}}} -
  {{{ g^{(1)}_{00}}(x,t)\,\frac{d^{2}g^{(1)}_{22}(x,t)}{dxdt}}\over
    {{{R(t)}^4}}} = 0
\end{eqnarray}
energy conservation:\\
\begin{eqnarray}
\label{ec,2}
&&{{-4\,{ g^{(2)}_{22}}(x,t)\,\dot{R}(t)}\over {{{R(t)}^3}}} -
  {{2\,{ g^{(2)}_{11}}(x,t)\,\dot{R}(t)}\over {{{R(t)}^3}}} +
  {{2\,\frac{dg^{(2)}_{22}(x,t)}{dt}}\over {{{R(t)}^2}}} +
  {{\frac{dg^{(2)}_{11}(x,t)}{dt}}\over {{{R(t)}^2}}} \nonumber\\
&&+
{{4\,{{{ g^{(1)}_{22}}(x,t)}^2}\,\dot{R}(t)}\over {{{R(t)}^5}}} +
  {{2\,{{{ g^{(1)}_{11}}(x,t)}^2}\,\dot{R}(t)}\over {{{R(t)}^5}}} -
  {{2\,{{{ g^{(1)}_{10}}(x,t)}^2}\,\dot{R}(t)}\over
{{{R(t)}^3}}}\nonumber\\
&& -
  {{4\,{ g^{(1)}_{00}}(x,t)\,{ g^{(1)}_{22}}(x,t)\,\dot{R}(t)}\over
    {{{R(t)}^3}}} - {{2\,{ g^{(1)}_{00}}(x,t)\,{ g^{(1)}_{11}}(x,t)\,
      \dot{R}(t)}\over {{{R(t)}^3}}}  -
  {{6\,{{{ g^{(1)}_{10}}(x,t)}^2}\,\dot{p}_{0}(t)\,\dot{R}(t)}\over
    {{{R(t)}^3}\,\dot{\rho}_{0}(t)}} \nonumber\\
&&+
  {{4\,{ g^{(1)}_{10}}(x,t)\,\frac{dg^{(1)}_{10}(x,t)}{dt}}\over
    {{{R(t)}^2}}} - {{2\,{ g^{(1)}_{22}}(x,t)\,
      \frac{dg^{(1)}_{22}(x,t)}{dt}}\over {{{R(t)}^4}}} +
  {{2\,{ g^{(1)}_{00}}(x,t)\,\frac{dg^{(1)}_{22}(x,t)}{dt}}\over
    {{{R(t)}^2}}} \nonumber\\
&& - {{{ g^{(1)}_{11}}(x,t)\,
      \frac{dg^{(1)}_{11}(x,t)}{dt}}\over {{{R(t)}^4}}} +
  {{{ g^{(1)}_{00}}(x,t)\,\frac{dg^{(1)}_{11}(x,t)}{dt}}\over
    {{{R(t)}^2}}} - {{{ g^{(1)}_{10}}(x,t)\,
      \frac{dg^{(1)}_{00}(x,t)}{dx}}\over {{{R(t)}^2}}} = 0
\end{eqnarray}
momentum conservation:\\
\begin{eqnarray}
\label{mc,2}
&&{{{ g^{(2)}_{10}}(x,t)\,\dot{p}_{0}(t)}\over {{{R(t)}^2}}} +
  {{{ p_{0}}(t)\,\frac{dg^{(2)}_{10}(x,t)}{dt}}\over
    {{{R(t)}^2}}} + {{{ \rho_{0}}(t)\,
      \frac{dg^{(2)}_{10}(x,t)}{dt}}\over {{{R(t)}^2}}} -
  {{{ p_{0}}(t)\,\frac{dg^{(2)}_{00}(x,t)}{dx}}\over
    {2\,{{R(t)}^2}}} - {{{ \rho_{0}}(t)\,
      \frac{dg^{(2)}_{00}(x,t)}{dx}}\over {2\,{{R(t)}^2}}}\nonumber\\
&&
-{{{ g^{(1)}_{10}}(x,t)\,{ g^{(1)}_{11}}(x,t)\,\dot{p}_{0}(t)}\over
     {{{R(t)}^4}}} + {{{ g^{(1)}_{00}}(x,t)\,{ g^{(1)}_{10}}(x,t)\,
      \dot{p}_{0}(t)}\over {{{R(t)}^2}}}+
  {{{ g^{(1)}_{10}}(x,t)\,{ p_{0}}(t)\,
      \frac{dg^{(1)}_{00}(x,t)}{dt}}\over {2\,{{R(t)}^2}}}\nonumber\\
&& +
  {{{ g^{(1)}_{10}}(x,t)\,{ \rho_{0}}(t)\,
      \frac{dg^{(1)}_{00}(x,t)}{dt}}\over {2\,{{R(t)}^2}}} -
  {{{ g^{(1)}_{11}}(x,t)\,{ p_{0}}(t)\,
      \frac{dg^{(1)}_{10}(x,t)}{dt}}\over {{{R(t)}^4}}} +
  {{{ g^{(1)}_{00}}(x,t)\,{ p_{0}}(t)\,
      \frac{dg^{(1)}_{10}(x,t)}{dt}}\over {{{R(t)}^2}}} \nonumber\\
&&-
  {{{ g^{(1)}_{11}}(x,t)\,{ \rho_{0}}(t)\,
      \frac{dg^{(1)}_{10}(x,t)}{dt}}\over {{{R(t)}^4}}} +
  {{{ g^{(1)}_{00}}(x,t)\,{ \rho_{0}}(t)\,
      \frac{dg^{(1)}_{10}(x,t)}{dt}}\over {{{R(t)}^2}}}\nonumber\\
&&  +
  {{{ g^{(1)}_{11}}(x,t)\,{ p_{0}}(t)\,
      \frac{dg^{(1)}_{00}(x,t)}{dx}}\over {2\,{{R(t)}^4}}} -
  {{{ g^{(1)}_{00}}(x,t)\,{ p_{0}}(t)\,
      \frac{dg^{(1)}_{00}(x,t)}{dx}}\over {2\,{{R(t)}^2}}} +
  {{{ g^{(1)}_{11}}(x,t)\,{ \rho_{0}}(t)\,
      \frac{dg^{(1)}_{00}(x,t)}{dx}}\over {2\,{{R(t)}^4}}}
\nonumber\\
&&-
  {{{ g^{(1)}_{00}}(x,t)\,{ \rho_{0}}(t)\,
      \frac{dg^{(1)}_{00}(x,t)}{dx}}\over {2\,{{R(t)}^2}}} = 0
\end{eqnarray}
Note that these equations transform into the system (\ref{eqs,n}) if
we insert in each order into the 00 component the energy conservation
equation, and if we divide the momentum conservation equation by
$(\rho_{0}+p_{0})$.
\end{appendix}

\end{document}